\documentclass{ws-ijmpa}

\usepackage{amssymb,amsmath}

\newcommand{\mbf}[1]{\boldsymbol{#1}}

\newcommand{\ee}[1]{{\rm e}^{#1}}
\newcommand{\ii}{{\rm i}}

\newcommand{\eq}{\begin{equation}}
\newcommand{\eqend}{\end{equation}}
\newcommand{\eqa}{\begin{eqnarray}}
\newcommand{\nonueqa}{\begin{eqnarray*}}
\newcommand{\eqaend}{\end{eqnarray}}
\newcommand{\nonueqaend}{\end{eqnarray*}}

\newcommand{\bma}[1]{\begin{array}{#1}}
\newcommand{\ema}{\end{array}}
\newcommand{\bc}{\begin{center}}
\newcommand{\ec}{\end{center}}

\newcommand{\real}{{\mathbb R}} 
\newcommand{\id}{{1\!\!1}} 

\newif\ifold             \oldtrue

\def\nn{\nonumber}

\newcommand{\Tr}[1]{\:{\rm Tr}\,#1}

\def\e{{\,\rm e}\,}

\def\be{\begin{equation}}
\def\ee{\end{equation}}
\def\bea{\begin{eqnarray}}
\def\eea{\end{eqnarray}}
\def\bd{\begin{displaymath}}
\def\ed{\end{displaymath}}

\def\dd{{\rm d}}

\def\ii{{\,{\rm i}\,}}

\newcommand{\beq}{\begin{eqnarray}}
\newcommand{\eeq}{\end{eqnarray}}

\makeatletter
\newdimen\normalarrayskip              
\newdimen\minarrayskip                 
\normalarrayskip\baselineskip
\minarrayskip\jot
\newif\ifold             \oldtrue            
\def\arraymode{\ifold\relax\else\displaystyle\fi} 
\def\@arrayskip{\ifold\baselineskip\z@\lineskip\z@
     \else
     \baselineskip\minarrayskip\lineskip2\minarrayskip\fi}
\def\@arrayclassz{\ifcase \@lastchclass \@acolampacol \or
\@ampacol \or \or \or \@addamp \or
   \@acolampacol \or \@firstampfalse \@acol \fi
\edef\@preamble{\@preamble
  \ifcase \@chnum
     \hfil$\relax\arraymode\@sharp$\hfil
     \or $\relax\arraymode\@sharp$\hfil
     \or \hfil$\relax\arraymode\@sharp$\fi}}
\def\@array[#1]#2{\setbox\@arstrutbox=\hbox{\vrule
     height\arraystretch \ht\strutbox
     depth\arraystretch \dp\strutbox
     width\z@}\@mkpream{#2}\edef\@preamble{\halign \noexpand\@halignto
\bgroup \tabskip\z@ \@arstrut \@preamble \tabskip\z@ \cr}%
\let\@startpbox\@@startpbox \let\@endpbox\@@endpbox
  \if #1t\vtop \else \if#1b\vbox \else \vcenter \fi\fi
  \bgroup \let\par\relax
  \let\@sharp##\let\protect\relax
  \@arrayskip\@preamble}
\makeatother

\begin{document}

\markboth{Richard J. Szabo}
{Magnetic Backgrounds and Noncommutative Field Theory}

%
\catchline{}{}{}{}{}
%

\title{MAGNETIC BACKGROUNDS\\ AND NONCOMMUTATIVE FIELD
  THEORY\footnote{Invited review to be published in {\sl International
  Journal of Modern Physics A.}}}

\author{\footnotesize RICHARD J. SZABO}

\address{Department of Mathematics, Heriot-Watt University\\
Scott Russell Building, Riccarton, Edinburgh EH14 4AS, U.K.}

\maketitle

\pub{{\tt HWM-04-01 , EMPG-04-01 , physics/0401142}}{January 2004}

\begin{abstract}

This paper is a rudimentary introduction, geared at non-specialists, to how
noncommutative field theories arise in physics and their applications
to string theory, particle physics and condensed matter systems.

\end{abstract}

\vskip 1cm

\noindent{\bf CONTENTS}

\begin{itemlist}
\item {\bf Introduction}
\item {\bf Strong Magnetic Fields}
  \begin{itemlist}
    \item The Landau problem
    \item The lowest Landau level
    \item Field theory
  \end{itemlist}
\item {\bf String Theory and D-Branes}
  \begin{itemlist}
    \item Noncommutative geometry in string theory
    \item D-branes
    \item String theory in magnetic fields
  \end{itemlist}
\item {\bf Noncommutative Quantum Field Theory}
  \begin{itemlist}
    \item Fundamental aspects
    \item UV/IR mixing
    \item Gauge interactions
    \item Violations of special relativity
  \end{itemlist}
\item {\bf The Fractional Quantum Hall Effect}
  \begin{itemlist}
    \item The Laughlin wavefunction
    \item Noncommutative Chern-Simons theory
  \end{itemlist}
\item {\bf Acknowledgments}
\item {\bf References}
\end{itemlist}

\section{Introduction \label{Intro}}

The idea behind spacetime noncommutativity is to replace the coordinates
$x^i$ of spacetime by Hermitian operators (also denoted
$x^i$) which obey the commutation relations
\beq
\left[x^i\,,\,x^j\right]=\ii\theta^{ij} \ ,
\label{spacetimenc}\eeq
where $\theta^{ij}$ is an antisymmetric tensor that may be constant, a
function of the coordinates $x^i$ themselves, or a function of both
coordinates and momenta. In the first instance the operators $x^i$
essentially define a Heisenberg algebra, while in the last case they
generate an algebra of pseudo-differential operators. This idea dates
back to the 1930's and is attributed to Heisenberg, who proposed it as a
means to control the ultraviolet divergences which plagued quantum
field theory. It was purported to ameliorate the problem of infinite
self-energies in a Lorentz-invariant way (for appropriate choices of
$\theta^{ij}$). The first phenomenological realization of this idea
took place not in particle physics but rather in condensed matter
physics by Peierls, who applied it to non-relativistic electronic systems in
external
magnetic fields (the celebrated Peierls substitution).\cite{Peierls1} The idea
propagated from Peierls onto Pauli, then onto Oppenheimer, who gave
the problem to his graduate student Snyder, leading to the first
published paper with systematic analysis on the subject in 1947.\cite{Snyder1}

A toy model of this realization comes about from taking $\theta^{ij}$
in (\ref{spacetimenc}) to be real-valued constants. In this case the
$\theta^{ij}$ play a role completely analogous to Planck's constant
$\hbar$ in the quantum phase space relation
$[x^i,p_j]=\ii\hbar\,\delta^i_{~j}$. In particular, there is a
spacetime uncertainty relation
\beq
\Delta x^i\,\Delta x^j\geq\mbox{$\frac12$}\,|\theta^{ij}| \ ,
\label{spacetimeuncert}\eeq
which implies that $|\theta^{ij}|$ measures the smallest patch of
observable area in the $(ij)$-plane. This gives a limit to the
resolution in which one may probe spacetime itself, and hence gives
insight into short-distance spacetime structure. The spacetime becomes
``fuzzy'' at very short distances, as there is no longer any definite
notion of a `point'. Such ideas are very common in models of quantum
gravity, which predicts that classical general relativity breaks down
at the Planck scale and requires a modification of the classical
notions of geometry. The recent surge of excitement in the subject has
come about from the discovery that such scenarios are realized explicitly in
string theory with D-branes.\cite{CDS1,DH1,LLS1,Schom1,SW1}

The purpose of this article is to provide a rudimentary exposition of
the interrelationships between the ideas of noncommutative geometry
that we have described above. The material is geared at the reader
with a reasonable background in theoretical physics, but no detailed prior
knowledge of noncommutative geometry or string theory. We will begin
by presenting a very simple quantum mechanical model, the Landau
problem,\cite{LL1}
which represents the simplest framework in which one can see noncommutative
field theory emerging as an effective description of the
dynamics~(Section~2). We will then briefly describe how this scenario
emerges in string theory~(Section~3), and how it leads to the study of
noncommutative quantum field theory~(Section~4). We also describe
various potential applications of this formalism to processes in
particle physics and in astrophysics. Finally, in Section~5 we
turn our attention back to the framework of Section~2 and describe a novel
application of noncommutative field theory in condensed matter physics
to the fractional quantum Hall effect.\cite{Suss1,Poly1,HMVR1} More
extensive reviews of noncommutative field theory may be found in
Refs.~\refcite{KS1,DN1,NCrevs}, where more complete lists of
references are also given.

\section{Strong Magnetic Fields}

In this Section we will describe how the fundamental notions of
noncommutative field theory arise in what is perhaps the simplest
possible physical setting, namely the quantum mechanics of the motion
of charged particles in two dimensions under the influence of a
constant, perpendicularly applied magnetic field.\cite{LL1} This introduces the
main technical points that the string theory inspirations do, as we
will describe in the next Section, but within a much simpler framework. It
also makes contact with the historical development of the subject
described in the previous Section and will be one of the motivations
for the application of noncommutative field theory that we describe in
Section~5. A similar introduction to noncommutativity is presented in
Ref.~\refcite{Jackiw1}.

\subsection{The Landau problem}

\begin{figure}
\centerline{\psfig{file=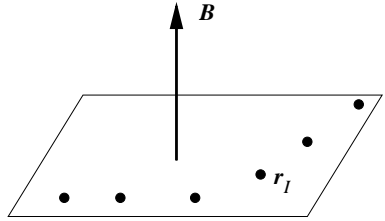,width=5cm}}
\vspace*{8pt}
\caption{Set-up for the Landau problem. A system of electrons
  moves in two-dimensions under the influence of an externally
  applied, constant perpendicular magnetic field.}
\label{Landau}\end{figure}

The Landau problem deals with a system of $N_e$ non-relativistic, interacting
electrons moving in two-dimensions. We denote their position
coordinates and velocities respectively by
\beq
{\mbf r}_I=\bigl(x_I\,,\,y_I\bigr)=\bigl(x_I^1\,,\,x_I^2\bigr) \ , ~~
{\mbf v}_I=\dot{\mbf r}_I \ ,
\label{mbfrv}\eeq
with $I=1,\dots,N_e$. The two-dimensional system is subjected to a
constant, external perpendicularly applied magnetic field ${\mbf
  B}=B\,\hat{z}$~(Fig.~\ref{Landau}). We will work in the gauge where the
corresponding vector potential is of the form
\beq
{\mbf A}(\mbf r_I)=(0,B\,x_I)
\label{vecpotB}\eeq
with ${\mbf B}={\mbf\nabla\mbf\times\mbf A}$. The Lagrangian governing this
motion is then given by
\beq
L=\sum_{I=1}^{N_e}\left(\frac{m_e}2\,{\mbf v}_I^2+\frac ec\,
{\mbf v}_I\cdot{\mbf A}({\mbf r}_I)-V({\mbf r}_I)\right)-
\sum_{I<J}U({\mbf r}_I-{\mbf r}_J) \ ,
\label{LandauLag}\eeq
where $V$ is the electron self-energy due its interaction, say, with
an impurity which is externally introduced into the system, and $U$ is a
pair-interaction potential between the electrons with the hard-core
condition $U(0)=0$.

Canonical quantization of this system proceeds in the usual way giving
the Hamiltonian operator
\beq
H=\sum_{I=1}^{N_e}\left(\frac{{\mbf\pi}_I^2}{2m_e}+V({\mbf r}_I)
\right)+\sum_{I<J}U({\mbf r}_I-{\mbf r}_J) \ ,
\label{LandauHam}\eeq
where
\beq
{\mbf\pi_I}=m_e\,{\mbf v}_I={\mbf p}_I-\mbox{$\frac ec$}\,
{\mbf A}({\mbf r}_I)
\label{kinmom}\eeq
is the (non-canonical) gauge-invariant kinematical momentum, while
${\mbf p}_I$ is the canonical momentum obeying the usual commutation
relations
\bea
\bigl[x_I\,,\,p_J^x\bigr]&=&\ii\hbar\,\delta_{IJ}~=~\bigl[y_I\,,\,
p_J^y\bigr] \ , \nn\\\bigl[x_I\,,\,y_J\bigr]&=&\bigl[p_I^x\,,\,
p_J^y\bigr]~=~0 \ ,
\label{cancomrels}\eea
and so on. From (\ref{vecpotB}) and (\ref{cancomrels}) it follows that
the components of the kinematical momentum (\ref{kinmom}) have the
non-vanishing quantum commutators
\beq
\bigl[\pi_I^x\,,\,\pi_J^y\bigr]=\ii\hbar\,\frac{eB}c\,\delta_{IJ} \ .
\label{kinmomcomrels}\eeq
Thus the physical (i.e. gauge invariant) momenta of the electrons in
the background magnetic field live in a noncommutative space.

The quantum momenta $\mbf\pi_I$ can be written in terms of harmonic oscillator
creation and annihilation operators. In the absence of interactions,
$V=U=0$, the energy eigenvalues of the normal-ordered Hamiltonian
(\ref{LandauHam}) are thus those of Landau levels
\beq
E=\sum_{I=1}^{N_e}\hbar\,\omega_c\,\left(n_I+\mbox{$\frac12$}\right) \
, ~~ n_I=0,1,2,\dots \ ,
\label{Landaulevels}\eeq
where
\beq
\omega_c=\frac{eB}{m_ec}
\label{cyclotron}\eeq
is the cyclotronic frequency of the classical electron orbits in the
magnetic field. The mass gap between Landau levels is the constant
$\Delta$ given by
\beq
\Delta=\mbox{$\frac12$}\,\hbar\,\omega_c \ .
\label{massgap}\eeq
In the next Subsection we will examine the Landau problem in the limit
whereby this mass gap becomes very large and all excited Landau levels
decouple from the ground state which has quantum numbers $n_I=0$ for
all $I=1,\dots,N_e$.

\subsection{The lowest Landau level}

In the previous Subsection we encountered a very simple situation in
which the {\it momentum} space of a physical system is
noncommutative. To see how a noncommutative {\it coordinate} space
arises, let us consider the strong field limit $B\to\infty$,
i.e. the energy regime $B\gg m_e$, or equivalently the (formal) limit
of small electron mass $m_e\to0$. In this limit the Lagrangian
(\ref{LandauLag}) reduces to
\beq
L~\longrightarrow~L_0=\sum_{I=1}^{N_e}\left(\frac{eB}c\,x_I\,
\dot y_I-V(x_I,y_I)\right)-\sum_{I<J}U({\mbf r}_I-{\mbf r}_J) \ .
\label{L0largeB}\eeq
For each $I=1,\dots,N_e$, this Lagrangian is of the form $p\,\dot
q-h(p,q)$, and so the coordinates $(\frac{eB}c\,x_I,y_I)$ form a
canonical pair giving
\beq
\bigl[x_I\,,\,y_I\bigr]=\ii\,\frac{\hbar\,c}{eB} \ .
\label{xycanpair}\eeq
These relations also follow formally from (\ref{kinmom}) and
(\ref{kinmomcomrels}) in the limit $B\to\infty$ with the symmetric
gauge choice $\mbf A(\mbf r_I)=\frac12\,(-B\,y_I,B\,x_I)$.

Let us now examine the precise meaning of the limit taken above. Since
the cyclotronic frequency (\ref{cyclotron}) diverges in the limit
$B\to\infty$ (or $m_e\to0$), the spacing (\ref{massgap}) between
Landau levels becomes infinite and the lowest $n_I=0$ level decouples
from all of the rest. Thus the strong field limit projects the quantum
mechanical system onto the lowest Landau level. This limit is in fact
a phase space reductive one. Since the reduced Lagrangian
(\ref{L0largeB}) is of first order in time derivatives, it effectively
turns the coordinate space into a phase space. In other words, the
original four-dimensional phase space (per electron) degenerates into
the two-dimensional configuration space. We conclude that
noncommuting coordinates arise in electronic systems constrained to
lie in the lowest Landau level.

We can write the commutation relations in the form introduced in the
previous Section as
\beq
\left[x_I^i\,,\,x_J^j\right]=\ii\,\delta_{IJ}\,\theta^{ij} \ ,
\label{deltathetanc}\eeq
where the noncommutativity parameters $\theta^{ij}$ are given by
\beq
\theta^{ij}=\frac{\hbar\,c}{eB}\,\epsilon^{ij}
\label{thetaB}\eeq
with $\epsilon^{ij}$ the antisymmetric tensor. The present context is
in fact the one in which the Peierls substitution was originally
carried out in 1933.\cite{Peierls1} If one introduces an impurity, described by
a
potential energy function $V$, into the electronic system as in
(\ref{L0largeB}), then one can compute the first order energy shift in
perturbation theory, due to the impurity, of the lowest Landau level by taking
the
components of the position coordinates $\mbf r_I=(x_I,y_I)$ in $V(\mbf
r_I)$ to be noncommuting variables.

Let us remark that one could have also arrived at this conclusion
within the Hamiltonian formalism. In the limit described above, the
Hamiltonian (\ref{LandauHam}) reduces to
\beq
H~\longrightarrow~H_0=\sum_{I=1}^{N_e}V(\mbf r_I)+\sum_{I<J}
U(\mbf r_I-\mbf r_J) \ .
\label{H0largeB}\eeq
This reduced Hamiltonian describes a topological theory, in that it
vanishes in the absence of the potentials, whereby there are no
propagating degrees of freedom. On the other hand,
the kinematical momenta (\ref{kinmom}) in this limit become
\beq
\mbf\pi_I=m_e\,\mbf v_I~\longrightarrow~\mbf0 \ ,
\label{pi0largeB}\eeq
and the condition $\mbf\pi_I\equiv\mbf0$ should be treated as constraints
on the theory. Since according to (\ref{kinmomcomrels}) they do not
commute, they are second class constraints in the usual Dirac classification of
constrained mechanical systems.\cite{Dirac1} This requires us to replace
canonical Poisson brackets with Dirac brackets, whose quantization
under the correspondence principle of quantum mechanics
gives the coordinate noncommutativity~(\ref{xycanpair}).\cite{Jackiw1}

\subsection{Field theory}

We now investigate the consequences of noncommutativity on second
quantization of the system, i.e. in its effective non-relativistic
field theory description. For this, we introduce the classical
electron density
\beq
\rho(\mbf r)=\sum_{I=1}^{N_e}\delta^2(\mbf r-\mbf r_I)
\label{edensity}\eeq
which defines the number operator for the many-body system with
$N_e=\int\dd^2\mbf r~\rho(\mbf r)$. Using it, we rewrite the
Hamiltonian (\ref{H0largeB}) in the lowest Landau level as
\beq
H_0=\int\dd^2\mbf r~\rho(\mbf r)\,V(\mbf r)+\frac12\,\int\!\!\int\dd^2\mbf r~
\dd^2\mbf r'~\rho(\mbf r)\,U(\mbf r-\mbf r'\,)\,\rho(\mbf r'\,) \ .
\label{H0rho}\eeq
The quantum density operator is defined in terms of electron creation
and annihilation operators $\psi^\dag(\mbf r)$ and $\psi(\mbf r)$ as
\beq
\rho(\mbf r)=\psi^\dag(\mbf r)\,\psi(\mbf r) \ .
\label{quantumrho}\eeq
However, it is difficult to define (\ref{edensity}) as a quantum
operator. We bypass this problem by working instead in momentum space
with the Fourier transform
\beq
\tilde\rho(\mbf k)=\int\dd^2\mbf r~\rho(\mbf r)~\e^{\ii\mbf k\cdot\mbf
  r} \ .
\label{rhoFourier}\eeq
Since $\mbf r_I$ is a noncommuting operator, we must specify an
ordering for (\ref{rhoFourier}). We shall use symmetric or Weyl
ordering defined by specifying the Fourier transform as
\beq
\tilde\rho(\mbf k)=\sum_{I=1}^{N_e}\e^{\ii\mbf k\cdot\mbf r_I} \ ,
\label{rhoWeyl}\eeq
which differs from normal ordering, say, by a momentum dependent phase
factor,
\beq
\tilde\rho(\mbf k)=\e^{\frac\ii2\,k_1k_2\,\theta^{12}}\,
\sum_{I=1}^{N_e}\e^{\ii k_1x_I}~\e^{\ii k_2y_I} \ .
\label{normordprod}\eeq

We can compute the commutation relations of the density operators
(\ref{rhoWeyl}) by using (\ref{deltathetanc}) and the
Baker-Campbell-Hausdorff formula to write
\beq
\e^{\ii\mbf k\cdot\mbf r_I}~\e^{\ii\mbf q\cdot\mbf r_I}=
\e^{-\frac\ii2\,\mbf k\times\mbf q}~\e^{\ii(\mbf k+\mbf q)\cdot\mbf
  r_I} \ ,
\label{BCHexp}\eeq
where we have defined the two-dimensional cross-product
\beq
\mbf k\times\mbf q=k_i\,\theta^{ij}\,q_j
\label{2Dcross}\eeq
and the noncommutativity parameters $\theta^{ij}$ are given by
(\ref{thetaB}). We thereby find that the operators (\ref{rhoWeyl})
close the trigonometric algebra\cite{FFZ1}
\beq
\bigl[\tilde\rho(\mbf k)\,,\,\tilde\rho(\mbf q)\bigr]=
2\ii\sin\left(\mbox{$\frac12$}\,\mbf k\times\mbf q\right)~
\tilde\rho(\mbf k+\mbf q) \ .
\label{rhoWeylcommrels}\eeq
This algebra coincides with the algebra of magnetic translation
operators for the fractional quantum Hall effect in the lowest Landau
level.\cite{Brown1,Zak1}

For an arbitrary c-number function $f(\mbf r)$ on the plane, we define
its classical average using the electron density as
\beq
\langle f\rangle=\int\dd^2\mbf r~\rho(\mbf r)\,f(\mbf r)=
\int\frac{\dd^2\mbf k}{(2\pi)^2}~\tilde\rho(\mbf k)\,
\tilde f(-\mbf k) \ .
\label{cnumavg}\eeq
In the quantum theory, we can compute the commutator of two such
averages by multiplying the trigonometric algebra
(\ref{rhoWeylcommrels}) on both sides by the convolution product
$\tilde f(-\mbf k)\,\tilde g(-\mbf q)$ of Fourier transforms, and then
integrate over the Fourier momenta to get
\beq
\bigl[\langle f\rangle\,,\,\langle g\rangle\bigr]=
\bigl\langle\,[f,g]_\star\bigr\rangle \ ,
\label{avgcomrels}\eeq
where we have introduced the star-commutator
\beq
[f,g]_\star(\mbf r)=(f\star g)(\mbf r)-(g\star f)(\mbf r) \ .
\label{starcommdef}\eeq
The function $f\star g$ is the noncommutative, associative
Gr\"onewold-Moyal star-product\cite{Gron1,Moyal1} of the functions $f$ and $g$
from the
theory of deformation quantization, and it may be expressed in
position space in terms of a non-local bi-differential operator as
\bea
(f\star g)(\mbf r)&=&f(\mbf r)~\exp\left(\mbox{$\frac\ii2$}\,
\overleftarrow{\partial_i}\,\theta^{ij}\,\overrightarrow{\partial_j}
\right)~g(\mbf r)\nonumber\\&=&f(\mbf r)\,g(\mbf r)+
\sum_{n=1}^\infty\frac{\ii^n}{2^n\,n!}\,\theta^{i_1j_1}\cdots
\theta^{i_nj_n}\,\partial_{i_1}\cdots\partial_{i_n}f(\mbf r)\,
\partial_{j_1}\cdots\partial_{j_n}g(\mbf r)\nonumber\\&&
\label{fstargdef}\eea
with $\partial_i=\partial/\partial x^i$. The relation
(\ref{avgcomrels}) thereby describes a very simple physical occurence
of the star-product for fields in a strong magnetic background. An
interpretation within noncommutative field theory of dipole behaviour
in a strong magnetic field, i.e. the Mott exciton, may also be
given.\cite{Rey1}

There are two important comments we should make about this
derivation. First of all, {\it only} the commutators of averages
coincide with star-commutators as in (\ref{avgcomrels}), and in
general one has
\beq
\langle f\rangle\langle g\rangle\neq\langle f\star g\rangle \ .
\label{avgneq}\eeq
Secondly, the expansion of the star-commutator (\ref{starcommdef}) for
$\theta^{ij}\to0$ (equivalently $B\to\infty$) yields, from
(\ref{thetaB}) and (\ref{fstargdef}), to lowest order the result
\beq
[f,g]_\star=\mbox{$\frac{\hbar\,c}{eB}$}\,\{f,g\}+O\left(
\mbox{$\frac1{B^2}$}\right) \ ,
\label{starcomPoisson}\eeq
where $\{f,g\}=\epsilon^{ij}\,\partial_if\,\partial_jg$ is the usual
Poisson bracket of the functions $f$ and $g$. This ``classical limit''
is the general foundation for the deformation quantization
programme,\cite{BFFLS1} in which the quantum phase space is
constructed by deforming the usual commutative product of functions on
classical phase space into a noncommutative star-product. The
star-commutator (\ref{starcommdef}) thereby encodes the usual
correspondence principle of quantum mechanics.

\section{String Theory and D-Branes}

In this Section we will describe a very precise realization of
spacetime noncommutativity which arises in string theory, and
which has sparked the enormous amount of activity in the subject over
the past few years. It is a direct generalization of the example
described in the previous Section. Its main virtue is that it
naturally induces what is known as a (relativistic) noncommutative field
theory, the
subject of the next Section. We will first describe heuristically why
noncommutative geometry is expected to play a role in string
theory,\cite{Witten1,LS1}
and then move our way towards a quantitative derivation of its
appearence. Unless explicitly written, in the remainder of this paper
we will assume natural units in which $\hbar=c=e=1$.

\subsection{Noncommutative geometry in string theory}

String theory is often regarded as the best candidate for a quantum
theory of gravitation, or more generally as a unified theory of all
the fundamental interactions. Within the framework of quantum gravity,
a noncommutative spacetime geometry is expected on quite general
grounds in any theory incorporating gravity into a quantum field
theory. At a semi-classical level, suppose we try to localize a
particle to within a Planck length $\lambda_{\rm P}\sim10^{-33}~{\rm
  cm}$ in any given plane of a spacetime. This would require that an
energy equal to the Planck mass $\sim10^{19}~{\rm GeV}/c^2$ must be
available to the particle. But such a process has enough energy to create a
black hole
and swallow the particle. We may avoid this paradox by requiring the
spacetime uncertainty principle\cite{DFR1}
\beq
\sum_{i<j}\Delta x^i\,\Delta x^j\geq\lambda_{\rm P}^2 \ .
\label{spacetimeuncert1}\eeq
This distorts the surrounding spacetime at very short distance scales
in the manner explained in Section~1. We conclude from this simple
analysis that spacetime noncommutativity is required when trying to
quantize the Einstein theory of general relativity.

\begin{figure}
\centerline{\psfig{file=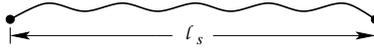,width=5cm}}
\vspace*{8pt}
\caption{The string length $\ell_s$. Strings alone cannot probe
  distances below their intrinsic size.}
\label{stringlength}\end{figure}

A similar scenario emerges directly from string theory. From the
analysis of ultra-high energy string scattering amplitudes,\cite{GM1} one is
led
to postulate the string-modified Heisenberg uncertainty
relation\cite{Ven1,ACV1}
\beq
\Delta x\geq\frac\hbar2\,\left(\frac1{\Delta p}+\ell_s^2\,\Delta
  p\right) \ ,
\label{stringmodHeisen}\eeq
where $\ell_s$ is the intrinsic string
length~(Fig.~\ref{stringlength}). This relationship reflects the
inherent non-locality of string theory, since it implies that the
extent of an object grows with its momentum. At large distance scales
$\gg\ell_s$
(formally the limit $\ell_s\to0$), wherein the strings effectively
look like point particles, it reduces to the standard phase space
relation in quantum mechanics with the spread decreasing with
momentum. Generically, by minimizing it with respect to $\Delta p$ one
finds that there is an absolute lower bound $\Delta x\geq(\Delta x)_{\rm min}$
on the measurability of distances in the spacetime given by the length
of the strings,
\beq
(\Delta x)_{\rm min}=\ell_s \ .
\label{Deltaxmin}\eeq
This simply means that strings cannot probe distances smaller than
their intrinsic size. String theory thereby requires a modification of
classical general relativity.

More generally, basic conformal symmetry arguments in string theory
lead to the anticipation of space/time uncertainty relations\cite{Yoneya1}
\beq
\Delta x\,\Delta t\geq\lambda_{\rm P}^2 \ .
\label{spacetimerels}\eeq
It is possible to realize such length scales using as probes not the
strings themselves, but rather certain non-perturbative open string degrees
of freedom known as {\it D-branes}.\cite{Horava1,DLP1,Polchinski1} In
fact, these objects allow one to probe even shorter, sub-Planckian
distance scales in string theory,\cite{DKPS1} and they enable
microscopic derivations of fairly generalized uncertainty relations
which include those described above as a subset.\cite{MS1}
They are therefore the natural degrees of freedom which capture
phenomena related to quantum gravitational fluctuations of the
spacetime. The beautiful aspect of this point of view is that these
phenomena can be treated systematically and at a completely
quantitative level in string theory.

\subsection{D-branes}

\begin{figure}
\centerline{\psfig{file=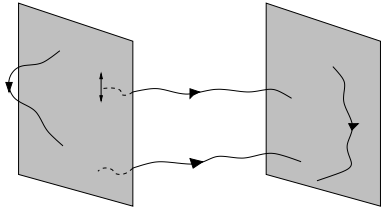,width=5cm}}
\vspace*{8pt}
\caption{A pair of D-branes with open string excitations which may
  start and end on the same brane, or stretch between the two of them.}
\label{DBranes}\end{figure}

Motivated by the discussion of the previous Subsection, let us now
systematically look at D-branes. A D-brane may be defined as a
hypersurface in spacetime onto which open strings attach (with
$D$irichlet boundary conditions). A schematic picture may be found in
Fig.~\ref{DBranes}. These degrees of freedom are actually {\it
  required} for the overall consistency of the string theory, which we
require to be unitary. The quantum theory of the open string
excitations induces a spectrum of fields which reside on the
branes. In the massless sector these include a gauge field $A_i$,
adjoint scalar fields $X^m$ describing the transverse fluctuations of the
D-branes
in spacetime, and fermion fields $\psi_\alpha$. Integrating out the massive
string modes on $N$ coincident D-branes leaves a low-energy effective
field theory which can be obtained as the dimensional reduction, to
the brane worldvolume, of ten-dimensional $U(N)$ Yang-Mills gauge
theory (more precisely, its supersymmetric extension).\cite{Witten2}

Let us study some features of this low-energy field theory
description. The reduction of the $F^2$ term of the Yang-Mills action
in ten spacetime dimensions (the critical superstring target space
dimension) leads to the Yang-Mills potential
\beq
V_{\rm YM}(X)=-\frac1{4g^2}\,\sum_{m\neq n}\Tr\bigl[X^m\,,\,X^n\bigr]^2 \ ,
\label{YMpot}\eeq
where $g$ is the Yang-Mills coupling constant and $X^m$ are $N\times
N$ Hermitian matrices. If $N=1$ then the $X^m$ correspond to the
fields which embed the D-brane into the Euclidean target spacetime. For $N>1$
they lose this geometric interpretation, and in this way the
appearence of noncommuting spacetime coordinates arises via a
dynamical mechanism.\cite{Witten2} The potential (\ref{YMpot}) is a sum of
non-negative terms, with $V_{\rm YM}(X)\geq0$ (note that
$\Tr[X^m,X^n]^2=-\Tr[X^m,X^n][X^m,X^n]^\dag$). Its global minimum
$V_{\rm YM}(X)=0$ is attained when the Hermitian matrices obey
\beq
\bigl[X^m\,,\,X^n\bigr]=0
\label{Xcomm0}\eeq
for each $m,n$. This means that the Hermitian matrices $X^m$ are
simultaneously diagonalizable in the vacuum state. Their simultaneous
real eigenvalues describe collective coordinates for the $N$ D-branes. Thus
the classical ground state corresponds to an ordinary classical
geometry. However, quantum fluctuations about the vacuum
(\ref{Xcomm0}) describe a spacetime with a noncommutative
geometry. The fluctuations correspond to turning on off-diagonal
matrix elements of the $X^m$'s and are associated with short open
string excitations between pairs of D-branes,\cite{Witten2} as depicted in
Fig.~\ref{DBranes}.

In this way the worldvolume field theories on the D-branes get altered
by quantum gravitational effects.\cite{DKPS1} To understand this modification,
it
is instructive to examine other classical vacua associated with the potential
(\ref{YMpot}). Generally, the equations of motion resulting from
variation of $V_{\rm YM}(X)$ are given by
\beq
\bigl[X_m\,,\,\bigl[X^m\,,\,X^n\bigr]\,\bigr]=0 \ .
\label{VXeoms}\eeq
A natural class of solutions is then provided by $X_0^m$ satisfying
the commutation relations
\beq
\bigl[X_0^m\,,\,X_0^n\bigr]=\ii\theta^{mn}
\label{X0commrels}\eeq
with $\theta^{mn}$ real-valued c-numbers, as in
(\ref{spacetimenc}). Taking the trace of both sides of
(\ref{X0commrels}) and using cyclicity shows that, for
$\theta^{mn}\neq0$, the relations (\ref{X0commrels}) can only be
satisfied by $N\times N$ matrices in the limit $N\to\infty$, i.e. by
operators acting on a separable Hilbert space which are not
trace-class, $\Tr(X_0^m)=\infty$. This is the usual situation for a
Heisenberg algebra. The expansion of the large $N$ matrices in
(\ref{YMpot}) about these more general classical vacua as
\beq
X^m=X_0^m+A^m(X_0)
\label{Xmexpand}\eeq
then determines a field theory for the $A^m$'s on a noncommutative
space.\cite{CDS1,AIIKKT1} This field theory in fact corresponds to the
noncommutative
Yang-Mills gauge theory which we will describe in the next
Section. The spacetime uncertainty relation (\ref{spacetimerels}) can
be seen to explicitly arise in this noncommutative gauge
theory.\cite{Rey2}

\subsection{String theory in magnetic fields}

We can make the appearence of noncommutative geometry in string theory
yet even more precise by considering the analog in string theory\cite{ACNY1,AMSS1} of
the Landau problem for strong magnetic fields that we studied in
detail in the previous Section. Let us consider the worldsheet field
theory for open strings attached to D-branes, which is defined by a
$\sigma$-model on the string worldsheet $\Sigma$ whose fields $y^i$
are maps from $\Sigma$ into the Euclidean target spacetime describing the
propagation of the strings. The geometry of the target space is
characterized by closed string supergravity fields, most notably the
spacetime metric $g_{ij}$ and the Neveu-Schwarz two-form $B_{ij}$
which we assume is non-degenerate. The action is
\beq
S_\Sigma=\frac1{4\pi\ell_s^2}\,\int\limits_\Sigma\dd^2\xi~
\Bigl(g_{ij}\,\partial^ay^i\,\partial_ay^j-2\pi\ii\ell_s^2\,B_{ij}\,
\epsilon^{ab}\,\partial_ay^i\,\partial_by^j\Bigr) \ ,
\label{SSigmadef}\eeq
where $\xi^a$, $a=1,2$ are local coordinates on the surface $\Sigma$ and
$\partial_a=\partial/\partial\xi^a$. The two-form $B_{ij}$ may be
regarded as a magnetic field on the D-branes.

In the case that the $B_{ij}$ are constant, the second term in
(\ref{SSigmadef}) is a total derivative which may be integrated by
parts to give the boundary action
\beq
S_{\partial\Sigma}=-\frac\ii2\,\oint\limits_{\partial\Sigma}
\dd t~B_{ij}\,y^i(t)\,\dot y^j(t) \ ,
\label{SbdrySigma}\eeq
where $t$ is the coordinate of the boundary $\partial\Sigma$ of the
string worldsheet residing on the D-brane worldvolume, and $\dot
y^i=\partial y^i/\partial t$. This is formally the same action that
arose in the Landau problem in the strong
field limit, and hence we can expect that quantization of the open
string endpoint coordinates $y^i(t)$ will induce a noncommutative
geometry on the D-brane. One needs to be somewhat careful though as
this is not a theory of particles. There is still the first term
present in the $\sigma$-model action (\ref{SSigmadef}) which
describes the dynamics of the interiors of the open strings, or
equivalently the closed string sector of the theory. It
reminds us that the point particles of (\ref{SbdrySigma}) are really
the endpoints of open strings. In particular, the two ends of an open
string couple directly to a background $B$-field and the string
becomes polarized as a dipole.

The remarkable observation is that there is a consistent low-energy
limit, called the Seiberg-Witten limit,\cite{SW1} which decouples all massive
string modes at the same time as scaling away the bulk part of the
string worldsheet dynamics from its boundary. It is defined by scaling
the spacetime metric as
\beq
g_{ij}\sim\ell_s^4\sim\varepsilon~\longrightarrow~0
\label{gijscale}\eeq
while keeping fixed the Neveu-Schwarz two-form field $B_{ij}$. Then
the worldsheet field theory is effectively described by the boundary
action (\ref{SbdrySigma}) alone and canonical quantization produces
the commutation relations
\beq
\left[y^i\,,\,y^j\right]=\ii\theta^{ij} \ , ~~ \theta=B^{-1}
\label{Dcommrels}\eeq
on $\partial\Sigma$. Thus the D-brane worldvolume becomes a
noncommutative space. Because of the point particle limit $\ell_s\to0$
taken in (\ref{gijscale}), the effective dynamics is governed in this
limit as usual by a field theory for the massless open string
modes. Following the analysis of the previous Section, we thus find
that the low-energy effective field theories on D-branes get modified
now to those defined with noncommuting coordinates, or equivalently by
star-products of the fields. In this way, string theory in the
presence of D-branes naturally leads to field theories on
noncommutative spaces. These models are the subject of the next
Section. It should be stressed that, as in the previous Section,
noncommutative field theories emerge here as {\it effective}
descriptions of the string dynamics. An equivalent description is
possible using ordinary theories on commutative spacetime.\cite{SW1}
However, the noncommutative setting is much more natural and both
conceptually and computationally useful, and it is from this formalism
that the true Planck scale physics of string theory may be captured by
quantum field theory.

\section{Noncommutative Quantum Field Theory}

One of the main interests in the emergence of field theories on
noncommutative spaces in the manner described above is that they
retain some of the non-locality of string theory, yet seem to be
well-defined as field theories. They thus present the remarkable
situation that many novel stringy features could be studied within the
simpler language of quantum field theory. To what extent this is true
is still to a large extent an unresolved issue. For instance, at
present, there still lacks a general, systematic renormalization programme for
handling such non-local field theories. These issues have addressed to
all orders of perturbation theory in
Refs.~\refcite{CR1,CR2,LSZ1,GW1}. Nevertheless, they {\it can} be
studied, and as field theories the non-locality gives them rather
exotic features which challenge the conventional wisdom of ordinary
quantum field theory. This is perhaps the greatest motivation for the
extensive study that they have seen, in that they are interesting on
their own as potentially well-defined examples of non-local field
theories. In this Section we shall briefly survey some of these
interesting new features, indicating how they capture some of the
non-locality of quantum gravity and highlighting some of the main
potential implications they could have on the structure of
spacetime. We will assume throughout that the noncommutative field
theories live on Euclidean spacetime. In Minkowski signature, making
time a noncommuting coordinate in the present context leads to severe
acausal effects and conceptual difficulties such as the precise
interpretation of Hamiltonian evolution. It also leads to violations
of unitarity and Lorentz invariance, as we discuss in Section~4.4. A
possible cure for this violation is suggested in Ref.~\refcite{Cure}
by integrating over all background fields corresponding to
noncommutativity parameters $\theta^{ij}$. This sum over backgrounds
is of course the natural recipe dictated by string theory and quantum
gravity, which are both unitary and covariant.

\subsection{Fundamental aspects}

For our purposes, we will define a noncommutative field theory as an
ordinary field theory whose commutative pointwise products of fields are
replaced with the noncommutative, associative star-product introduced
in Section~2, i.e. we replace
\beq
f(x)\,g(x)~\longmapsto~(f\star g)(x)=f(x)~\exp
\left(\mbox{$\frac\ii2$}\,\overleftarrow{\partial_i}\,\theta^{ij}\,
\overrightarrow{\partial_j}\right)~g(x) \ ,
\label{fgfstarg}\eeq
with $(\theta^{ij})$ an invertible antisymmetric matrix.
With respect to this product, an elementary computation shows that the
spacetime coordinates $x=(x^i)$ obey the required commutation
relations (\ref{spacetimenc}),
\beq
\left[x^i\,,\,x^j\right]_\star=x^i\star x^j-x^j\star
x^i=\ii\theta^{ij} \ .
\label{spacetimencstar}\eeq
Under Fourier transformation of fields, the star-product
(\ref{fgfstarg}) corresponds to the modification of the usual Fourier
convolution product as
\beq
\tilde f(k)\,\tilde g(q)~\longmapsto~\tilde f(k)\,\tilde g(q)~
\e^{\ii k\times q} \ , ~~ k\times q=k_i\,\theta^{ij}\,q_j \ ,
\label{starconvprod}\eeq
where the tildes denote Fourier transforms. The alteration
(\ref{starconvprod}) in momentum space exemplifies the inherent
non-locality of the star-product of fields. If $\theta$ is the average
magnitude
of a matrix element of $(\theta^{ij})$, then $1/\sqrt\theta$ is the energy
threshold beyond which a particle moves and interacts in a distorted
spacetime. The product deformation thus becomes effective at energies
$E$ with $E\,\sqrt\theta\ll1$, wherein not only are the interactions
between the fields significantly modified, but so are the quanta which
mediate these interactions.

As an explicit example, let us consider the noncommutative $\phi_4^4$
theory which is defined by the Euclidean action
\beq
S_\phi=\int\dd^4x~\left[\frac12\,(\partial_i\phi)^2+\frac{m^2}2\,\phi^2+
\frac\lambda{4!}\,\phi\star\phi\star\phi\star\phi\right] \ ,
\label{phi44theory}\eeq
where $\phi(x)$ is a real scalar field on $\real^4$. Note that only
interactions are modified by noncommutativity. The free field theory
is unaffected owing to the fact that the spacetime average of the
product of two fields is unchanged by the deformation,
\beq
\int\dd^4x~(f\star g)(x)=\int\dd^4x~f(x)\,g(x) \ ,
\label{intfstargunchanged}\eeq
which is easily derived from the representation (\ref{fgfstarg}) via
an integration by parts (assuming appropriate decay behaviour of the
fields at infinity in $\real^4$). In scalar field theories such as this one,
by using (\ref{starconvprod}) one can easily compute interaction vertices
in momentum space as
\unitlength=1.00mm
\linethickness{0.4pt}
\begin{equation}
\begin{picture}(100.00,25.00)
\thinlines
\put(32.00,12.00){\line(-1,1){10.00}}
\put(32.00,12.00){\line(-1,-1){10.00}}
\put(32.00,12.00){\circle*{1.50}}
\put(18.00,22.00){\makebox(0,0)[l]{$k_1$}}
\put(18.00,2.00){\makebox(0,0)[l]{$k_2$}}
\put(26.00,12.00){\makebox(0,0)[l]{$\lambda$}}
\put(32.00,12.00){\line(1,1){10.00}}
\put(32.00,12.00){\line(1,-1){10.00}}
\put(40.00,12.00){\makebox(0,0)[l]{$\vdots$}}
\put(44.00,2.00){\makebox(0,0)[l]{$k_3$}}
\put(44.00,22.00){\makebox(0,0)[l]{$k_n$}}
\put(48.00,12.00){\makebox(0,0)[l]{$~=~\lambda~
\e^{\frac\ii2\,\sum\limits_{I<J}k_I\times k_J}$}}
\end{picture}
\label{NCvertex}\end{equation}
along with the usual momentum conservation constraint
\beq
k_1+k_2+\ldots+k_n=0 \ .
\label{momcons}\eeq
We see that the noncommutative vertex (\ref{NCvertex}) is momentum
dependent, in contrast to the usual case whereby the tree-level vertex
function would simply be equal to the coupling constant $\lambda$. In
particular, this modifed interaction vertex is only cyclically
invariant under permutations of the momenta $k_I$, and so one needs to
keep careful track of the order of the momenta flowing into the vertex, again
in contrast to what would occur in ordinary scalar field theory.

It is well-known from multi-colour gauge theories and matrix models how
to keep track of the cyclic ordering.\cite{tHooft1,BIZ1} One doubles each line
of a
Feynman graph into ribbons. The ribbon graphs now have topology associated to
them, and one can characterize them into two sets, called planar and
non-planar.\cite{Filk1,IIKK1,MVRS1} The planar diagrams are those which can be
drawn on the
surface of a sphere or the plane without crossing any of the
ribbons. They are dynamically characterized by the fact that the totality
of the momenta of internal line contractions vanishes. In this case
the noncommutative phase factor (\ref{NCvertex}) does not
significantly alter the analytic expression for the corresponding amplitude. It
is
equal to the ordinary, $\theta=0$ commutative diagram, times a possible
$\theta$-dependent phase factor coming from the external momenta of
the graph. In particular, there is no change in the convergence
properties as compared to the commutative case. Much more interesting
are the non-planar diagrams, those which cannot be drawn on the
surface of a sphere or the plane. Dynamically, non-trivial internal
momentum contractions remain, and these graphs typically modify the
ultraviolet behaviour in a significant way.\cite{MVRS1} Naively, the rapid
phase
oscillations of (\ref{NCvertex}) suppress large momentum modes and
would appear to make the amplitude ultraviolet finite. As we will
discuss in the next Subsection, this is a subtle point, because the
phase factors (\ref{NCvertex}) become ineffective at vanishing
momenta, or equivalently the commutative ultraviolet divergence must
reappear at $\theta=0$ and the amplitude exhibits non-analytic
behaviour as a function of the noncommutativity parameter. This is a
surprising feature of the quantum field theory, as the classical field
theory smoothly reduces to its commutative counterpart at $\theta=0$.

\subsection{UV/IR mixing}

Let us now explore in a bit more detail the non-locality induced by
noncommutativity, and in particular the convergence properties of
Feynman diagrams in the quantum field theory. At a semi-classical
level, the non-locality of the star-product (\ref{fgfstarg}) itself
already produces exotic effects. Suppose that $f$ and $g$ are
wavepackets which are supported in a region of small size
$\Delta\ll\sqrt\theta$. One can then show, essentially from the
momentum representation of the star-product, that $f\star g$ is
non-zero in a large region of size $\theta/\Delta$. An extreme example
of this non-locality is provided by the star-product of two
infinitely-localized delta-functions, which is constant throughout
space,
\beq
\delta(x)\star\delta(x)=\frac1{\bigl|\det(\pi\,\theta)\bigr|} \ .
\label{deltastar}\eeq
In other words, the behaviour of the field theory at very short
distances, where the effects of spacetime noncommutativity are
significant, influences its long wavelength properties.

The effect just described has rather profound consequences in the
quantum field theory, and it leads to the famous {\it UV/IR mixing}
property of noncommutative field theories.\cite{MVRS1} If a Feynman diagram
requires an ultraviolet cutoff $\Lambda$ to regularize it, then this
automatically induces an effective infrared cutoff
\beq
\Lambda_0=\frac1{\theta\,\Lambda}
\label{IReff}\eeq
on the graph. We will describe below some of the remarkable consequences
of this mixing of energy scales, but let us first point out a simple
physical picture of this effect. As we did in (\ref{BCHexp}), from the
Baker-Campbell-Hausdorff formula and the commutation relation
(\ref{spacetimencstar}), one can straightforwardly compute
\beq
\e^{\ii k\cdot x}\star\e^{\ii q\cdot x}\star\e^{-\ii k\cdot x}
=\e^{-\frac\ii2\,k\times q}~\e^{\ii(k+q)\cdot x}\star\e^{-\ii k\cdot
  x}=\e^{\ii q_i(x^i-\theta^{ij}\,k_j)} \ .
\label{starBCH}\eeq
By Fourier transformation, it then follows that star-conjugation of
fields by plane waves induces a non-local spacetime translation as
\beq
\e^{\ii k\cdot x}\star f(x^i)\star\e^{-\ii k\cdot x}=f(x^i-\theta^{ij}\,
k_j) \ .
\label{starconjwaves}\eeq
We interpret (\ref{starconjwaves}) to mean that the quanta in
noncommutative field theory include ``dipoles'',\cite{Sheikh1,BS1} i.e.
extended, rigid
objects whose length or electric dipole moment $\Delta x^i$ grows with
its center of mass momentum $p_j$,
\beq
\Delta x^i=\theta^{ij}\,p_j \ .
\label{dipolerel}\eeq
These quanta are responsible for many of the stringy effects that
noncommutative field theories exhibit (c.f.~(\ref{stringmodHeisen})),
and they are just like the electron-hole bound states which arise in a
strong magnetic field. The dipoles interact by joining at their ends,
and this gives a simple picture of the non-local nature of the
interactions in noncommutative quantum field theory.

Let now examine the UV/IR mixing property at a more quantitative
level. Consider again, for definiteness, the noncommutative $\phi_4^4$
field theory with action (\ref{phi44theory}). Using the steps
described in the previous Subsection, one can work out the
one-particle irreducible effective action at one-loop order in
momentum space as\cite{MVRS1}
\beq
S_{\rm 1PI}=\int\frac{\dd^4k}{(2\pi)^4}~\frac12\,\tilde\phi(k)\,
\tilde\phi(-k)\,\left[k^2+\tilde m^2+\lambda\,\frac{\Lambda_{\rm
      eff}^2}{96\pi^2}-\lambda\,\frac{\Lambda^2}{96\pi^2}\,
\ln\left(\frac{\Lambda_{\rm eff}^2}{\Lambda^2}\right)\right] \ ,
\label{1PIeffaction}\eeq
where
\beq
\tilde m^2=m^2+\lambda\,\frac{\Lambda^2}{48\pi^2}-\lambda\,
\frac{\Lambda^2}{48\pi^2}\,\ln\left(\frac{\Lambda^2}{m^2}\right)
\label{tildem}\eeq
is the usual $\phi_4^4$ mass renormalization at one-loop order, and
\beq
\Lambda_{\rm eff}^2=\frac1{\frac1{\Lambda^2}+k_i\,\left(\theta^2
\right)^{ij}\,k_j}
\label{Lambdaeff}\eeq
is the momentum-dependent effective ultraviolet cutoff. From these
expressions one clearly sees that the limits $\theta\to0$ or $k\to0$
(the infrared limit) and $\Lambda\to\infty$ (the ultraviolet limit) do
not commute. Taking $\Lambda\to\infty$ leaves infrared singularities
as $k\to0$, as then the noncommutative phase factors (\ref{NCvertex}) become
ineffective at damping the ultraviolet behaviour in this
momentum range. This feature makes standard Wilsonian renormalization
treacherous, as it would normally require a clear separation of high
and low momentum scales. The low-energy effective field theory here does {\it
  not} decouple from the high-energy dynamics. The higher the cutoff
$\Lambda$ is, the more sensitive are the amplitudes to the lowest
energies available.

\subsection{Gauge interactions}

Let us now study the example of noncommutative gauge theory which can
be expected to tie in to the properties of our observable world, and
which is also inspired by the string theory applications that we
described in the last Section.\cite{CDS1,LLS1,SW1} The noncommutative
Yang-Mills action
for a $U(N)$ gauge field $A_i(x)$ on $\real^4$ is given by
\beq
S_{\rm NCYM}=-\frac1{4g^2}\,\int\dd^4x~\Tr\,F_{ij}(x)^2 \ ,
\label{SNCYM}\eeq
where
\beq
F_{ij}=\partial_iA_j-\partial_jA_i-\ii[A_i,A_j]_\star
\label{NCfieldstrength}\eeq
is the noncommutative field strength tensor. The curvature
(\ref{NCfieldstrength}) is a non-local deformation of the usual $U(N)$
field strength
\beq
F_{ij}=\partial_iA_j-\partial_jA_i-\ii[A_i,A_j]+O\bigl(\theta,(\partial
A)^2\bigr) \ .
\label{fieldstrengthdeform}\eeq
The action (\ref{SNCYM}) is invariant under the noncommutative version
of the usual gauge transformations, leading to the star-gauge
invariance
\beq
A_i~\longmapsto~U\star A_i\star U^{-1}+\ii U\star\partial_iU^{-1} \ ,
\label{stargaugeinv}\eeq
where $U(x)$ is an $N\times N$ star-unitary matrix field on $\real^4$,
\beq
U\star U^\dag=U^\dag\star U=\id \ .
\label{starunitary}\eeq

The presence of the star-product in the gauge transformation rule
(\ref{stargaugeinv}) mixes spacetime and colour degrees of freedom in
an intricate way. In fact, noncommutative gauge transformations in a
certain sense generate the infinite unitary group $U(\infty)$.\cite{LSZ2} This
implies that star-gauge invariance will contain geometrical symmetries
of the spacetime, in particular the symplectomorphisms of $\real^4$
with respect to the Poisson bi-vector
$\theta^{ij}\,\frac\partial{\partial x^i}\otimes\frac\partial{\partial
  x^j}$. To understand this point, let us consider the basic
plane wave fields in the case $N=1$,
\beq
U_a(x)=\e^{\ii(\theta^{-1})_{ij}\,a^j\,x^i} \ .
\label{Uaxdef}\eeq
{}From (\ref{starBCH}) one easily checks that they are star-unitary,
\beq
U^{~}_a\star U_a^\dag=U^\dag_a\star U_a^{~}=1 \ ,
\label{Uastarunitary}\eeq
while from (\ref{starconjwaves}) it follows that they implement
translations of fields by the vector $a=(a^i)\in\real^4$,
\beq
U^{~}_a(x)\star g(x)\star U_a^\dag(x)=g(x+a) \ .
\label{Uaxtransl}\eeq
{}From (\ref{stargaugeinv}) and (\ref{Uaxdef}) one then finds that the
corresponding star-gauge transformations are given by
\beq
A_i(x)~\longmapsto~A_i(x+a)-\left(\theta^{-1}\right)_{ij}\,a^j \ .
\label{stargaugetransl}\eeq
Up to a global translation of $A_i$, which leaves invariant the field
strength tensor (\ref{NCfieldstrength}), we see that spacetime
translations are equivalent to gauge transformations in noncommutative
Yang-Mills theory.\cite{GHI1} The only other known theory with such a
geometrical
gauge symmetry is gravity, and we may thereby conclude that
noncommutative gauge theory provides a toy model of general
relativity.

This translational symmetry can be naturally gauged within this
framework, and the noncommutative gauge theory can be thereby shown to
reduce to a teleparallel formalism of general
relativity.\cite{LS2} This feature fits in well with the hope that
noncommutative gauge theories capture important stringy
properties.\cite{LS3,IIKK2} A particularly important consequence of
this spacetime-colour mixing is that there are no {\it local}
gauge-invariant operators in noncommutative Yang-Mills
theory.\cite{IIKK1,MVRS1,GHI1} From the appropriate analogs of Wilson
line operators, it is in fact possible to derive closed string,
gravitational degrees of freedom from these open string noncommutative
gauge theories.\cite{OO1,DK1} Various aspects concerning the
adequacy of open Wilson lines for closed string like behaviour in
generic noncommutative field theories have been described in detail
in~\refcite{Rey3,Rey4,Rey5}.

\subsection{Violations of special relativity}

In this Subsection we will give an overview of some of the broad
phenomenological applications that spacetime noncommutativity may
have. Let us first observe that noncommutative field theories violate
Lorentz invariance. In the string picture, this violation is due to
the expectation value of the supergravity field $B_{ij}$. Focusing on
the four-dimensional situation, the noncommutativity tensor
$\theta^{ij}$ provides a directionality
$(\mbf\theta)_i=\epsilon_{ijk}\,\theta^{jk}$ in space, within any {\it
  fixed} inertial frame. Thus noncommutative field theory is not
invariant under rotations or boosts of localized field configurations
within a fixed observer inertial frame of reference.

The orientation $\mbf\theta$ can be used to provide stringent constraints
on the observable magnitude of the noncommutativity parameters
$\theta^{ij}$. Let us briefly summarize a few of the analyses that
have been made:
\begin{romanlist}[(iii)]
\item One can compare the noncommutative extension of the standard
  model with certain Lorentz-violating extensions. Noncommutative
  field theories are CPT symmetric,\cite{Sheikh2} hence so should be these
  commutative extensions. Comparison with the known literature can be used to
  derive the bound\cite{CHKLO1}
\beq
\theta<(10~{\rm TeV})^{-2}
\label{thetabound}\eeq
by using atomic clock comparison studies and a model for the ${}^9$Be
nucleus wavefunction.
\item We can also compare noncommutative quantum electrodynamics with
  some of the more standard QED processes, by taking into account the
  motion of the laboratory frame relative to $\mbf\theta$. Among the
  many scenarios considered have been high-energy $e^+e^-$ and hadron
  scattering, CP-violation, the anomalous magnetic moment $(g-2)\mu$,
  and so on. A review of these phenomenological applications is found
  in Ref.~\refcite{HK1}, where a complete list of references is also
  given.
\item Finally, we can compare noncommutative QED with low-energy
  atomic transitions. For example, in the noncommutative version of
  the Lamb shift in hydrogen,\cite{CS-JT1} the leading modification of the
Coulomb
  potential is given by
\beq
V_{\rm C}(\mbf r)=-\frac{e^2}r-\frac{e^2\,(\mbf r\mbf\times\mbf
k)\cdot\mbf\theta}
{r^3}+O\left(\theta^2\right) \ .
\label{Lambshift}\eeq
A recent overview of the various bounds obtained on spacetime
noncommutativity is presented in Ref.~\refcite{Calmet1}.
\end{romanlist}

Let us now turn to the phenomenological implications of UV/IR
mixing. From (\ref{1PIeffaction})--(\ref{Lambdaeff}) we see that
noncommutativity modifies the standard dispersion relation of special
relativity to
\beq
E^2=\mbf k^2+m^2+\Delta M^2\left(\frac1{k\,\theta}\right) \ ,
\label{moddispersion}\eeq
where $\Delta M^2(\Lambda)$ is the ultraviolet divergent mass
renormalization. The deformation $\Delta M^2$ in (\ref{moddispersion})
induces a violation of classical special relativity. We can compare
this formula to experimental measurements in the energy range
\beq
\Lambda_0<E<\Lambda=\frac1{\theta\,\Lambda_0} \ ,
\label{energyrange}\eeq
where $\Lambda_0$ is an experimentally determined phenomenological
infrared scale. This implies that one can only compare the effects of
UV/IR mixing with very high energy experimental data.

For example, the photon dispersion relation in
noncommutative electrodynamics is given by\cite{MichMor}
\beq
\omega=c\,k-c\,k\,\mbf\theta_\perp\cdot\mbf B_\perp \ ,
\label{photondisprel}\eeq
where $\mbf\theta_\perp,\mbf B_\perp$ are the components of
$\mbf\theta$ and a constant background magnetic induction $\mbf B$
transverse to the direction of light propagation $\mbf k$, i.e. $\mbf
k\cdot\mbf\theta_\perp=\mbf k\cdot\mbf B_\perp=0$. To reproduce the
bound (\ref{thetabound}), one would need to arrange a Michelson-Morley
type interferometry experiment with visible light, i.e. $B$ of the
order of a Tesla, and with leg lengths $l_1,l_2$ obeying
$l_1+l_2\geq10^{18}~{\rm cm}$, which is of the order of a parsec. This
is probably impractical for galactic magnetic
fields.\cite{Jackiw1,MichMor}

Finally, various cosmological comparisons can be made based on the
deformed dispersion relation (\ref{moddispersion}).\cite{A-CDNS1} For example,
in
certain models of astrophysical gamma-ray bursts, spacetime foam
induces dispersion. This comes about from ultra-high energy cosmic ray
thresholds (the GZK cutoff) on cosmic proton energies due to the
photopion production reaction $p+\gamma\longrightarrow p+\pi$ with cosmic
microwave
background radiation. Relations such as (\ref{moddispersion}) can be
used to explain the various puzzling and paradoxical observations of
cosmic rays.

\section{The Fractional Quantum Hall Effect}

Having dispelled with our tour of the significance of noncommutative
field theory in high-energy physics, we will now go back to our
motivating example of Section~2 and apply what we have learnt about
these novel field theories. A very precise application of
noncommutative field theory is in fact to the fractional quantum Hall
effect. A particular such model provides a mean field theory
description that is far superior to its commutative version and which
displays the correct quantitative features expected from condensed
matter physics.

\subsection{The Laughlin wavefunction}

Let us begin with a brief review of some basic and well-known facts
about the mean field theory of the Landau problem.\cite{Girvin1} In the
fractional
quantum Hall effect, the interactions lead to a state similar to the
filled lowest Landau level, but allowing for fractionally charged
quasi-particle excitations. Let $m$ be a positive integer. The ratio
of the electron density to the density of the lowest Landau level is
the filling fraction $\nu$, and at $\nu=\frac1m$ a good microscopic
description of such a state is provided by the $N_e$-electron Laughlin
wavefunction\cite{Laughlin1}
\beq
\Psi_{1/m}(z_1,\dots,z_{N_e})=\prod_{I<J}\left(z_I-z_J\right)^m~
\e^{-\frac1{2\theta}\,\sum\limits_{I=1}^{N_e}|z_I|^2} \ ,
\label{Psi1m}\eeq
where $\theta=\hbar\,c/eB$ and $z_I=x_I+\ii y_I$ are complex
coordinates on the plane for each $I=1,\dots,N_e$. In canonical
quantization, the pairs $(z_I,\overline{z_I}\,)$ define essentially
$N_e$ decoupled harmonic oscillators. The state (\ref{Psi1m}) has
charge density equal to $\frac1m$ times the density of a filled Landau
level.

A quasi-particle at the point $z_0$ is created from the state
(\ref{Psi1m}) by acting with the operator
\beq
{\cal Q}(z_0)=\prod_{I=1}^{N_e}\left(\,\overline{z_I}-\overline{z_0}\,
\right) \ ,
\label{quasipartop}\eeq
where we represent the oscillator algebra by
$\overline{z_I}=2\theta\,\frac\partial{\partial z_I}$. The
quasi-particle states are characterized by the two properties they
have:
\begin{romanlist}[(ii)]
\item Fractional charge $\frac1m$.
\item Fractional exchange statistics, i.e. a $2\pi$ rotation of the
  relative coordinate of a two-quasi-particle state multiplies the
  state by the phase factor $\e^{2\pi\ii/m}$.
\end{romanlist}

The low-energy excitations of the ground state may be described by a
Landau-Ginzburg theory of a superfluid density $\phi$ minimally coupled to a
fictitious abelian vector potential $A_i$ in $2+1$ dimensions.\cite{Fradkin1}
The
original quasi-particles are magnetic vortex solutions of this model, while
their fractional
statistics is reproduced by including in the action an abelian
Chern-Simons term for the gauge potential,
\beq
S_{\rm CS}=\frac{\ii m}{2\pi}\,\int\dd^3x~\epsilon^{ijk}\,A_i\,
\partial_jA_k \ .
\label{SCSaction}\eeq
The Gauss law for this gauge-matter coupled theory implies that a
vortex of unit magnetic charge also carries electric charge
$\frac1m$. The Aharonov-Bohm phase factors about the magnetic vortex
then lead to the appropriate fractional statistics. This model
effectively describes the Landau problem as a quantum Hall {\it fluid}
in terms of a hydrodynamical gauge theory.\cite{BS2,FJL1}

\subsection{Noncommutative Chern-Simons theory}

We will now demonstrate that the noncommutative version of the
Chern-Simons action (\ref{SCSaction}) leads directly to a very
efficient description of the quasi-particle excitations,\cite{Suss1}
in which the elevation of the hydrodynamic gauge theory to a
noncommutative gauge theory captures the graininess of the quantum
Hall fluid. The primary motivation {\it a priori} for making the spatial
directions $\mbf
r=(x^1,x^2)$ noncommuting variables resides in our analysis of Section~2. The
time coordinate $x^0=t$ is left as an ordinary commutative
variable. The action is defined by
\beq
S_{\rm NCS}=\frac{\ii m}{2\pi}\,\int\dd^3x~\epsilon^{ijk}\,
\left(A_i\,\partial_jA_k+\mbox{$\frac23$}\,A_i\star A_j
\star A_k\right) \ .
\label{SNCSdef}\eeq
It is invariant under the usual noncommutative gauge transformations
$U$ in (\ref{stargaugeinv},\ref{starunitary}) which are trivial at
spatial infinity,\cite{CS1} i.e. $U(t,\mbf r)\to\id$ at $|\mbf r|\to\infty$. In
the temporal gauge $A_0=0$, the action (\ref{SNCSdef}) is formally the
same as its commutative counterpart~(\ref{SCSaction}), i.e.
\beq
S_{\rm NCS}[A_0=0]=\frac{\ii m}{2\pi}\,\int\dd
t~\int\dd^2\mbf r~\epsilon^{0ij}\,A_i\,\partial_tA_j \ .
\label{SNCStemporal}\eeq
However, now the Gauss law, i.e. the equation of motion for $A_0$,
involves the {\it noncommutative} field strength tensor and is given
by
\beq
F_{ij}=\partial_iA_j-\partial_jA_i-\ii[A_i,A_j]_\star=0 \ .
\label{NCGausslaw}\eeq

The crucial observation now is that the action and constraint arise
from a matrix model in $0+1$ dimensions with action\cite{Suss1,Poly1}
\beq
S_{\rm MCS}=\frac{2m}\theta\,\int\dd t~\Tr\left(
\mbox{$\frac12$}\,\epsilon^{ij}\,X_i\,D_tX_j+\theta\,A_0\right) \ ,
\label{SMCSdef}\eeq
where $X^i$, $i=1,2$ and $A_0$ are $N\times N$ time-dependent Hermitian
matrices, and $D_t=\partial_t-\ii A_0$. This is a gauged $U(N)$ matrix
quantum mechanics which we will call {\it matrix Chern-Simons
  theory}. To establish its equivalence with the noncommutative gauge
theory defined by (\ref{SNCSdef}), we write (\ref{SMCSdef}) in the
$A_0=0$ gauge to get
\beq
S_{\rm MCS}[A_0=0]=\frac m\theta\,\int\dd t~\epsilon^{ij}\,
X_i\,\partial_tX_j \ ,
\label{SMCStemporal}\eeq
and note that the constraint arising from varying the action
(\ref{SMCSdef}) with respect to the non-dynamical variable $A_0$ is
given by the commutation relation
\beq
\left[X^1\,,\,X^2\right]=\ii\theta \ .
\label{A0constr}\eeq
In particular, as we discussed earlier, the $X^i$ are necessarily
infinite-dimensional matrices with $N\to\infty$, i.e. operators acting
on a separable Hilbert space.

Let us now expand the action (\ref{SMCStemporal}) and constraint
(\ref{A0constr}) about a particular time-independent solution $y^i$,
as we did in (\ref{Xmexpand}), i.e. we write
\beq
X^i=y^i+\theta\,\epsilon^{ij}\,A_j
\label{Xyexpand}\eeq
with
\beq
\left[y^i\,,\,y^j\right]=\ii\theta\,\epsilon^{ij}
\label{ycommrels}\eeq
and $A_i$ functions of the noncommuting coordinates $y^i$. From
(\ref{ycommrels}) it follows that the $y^i$'s act on functions of $y$
alone as
\beq
\bigl[y^i\,,\,f(y)\bigr]=\ii\theta\,\epsilon^{ij}\,\frac{\partial
  f(y)}{\partial y^j} \ .
\label{yiderivact}\eeq
We then use the usual Weyl-Wigner correspondence of noncommutative
field theory,\cite{NCrevs} which generally reflects the fact that
noncommutative
fields are most naturally thought of as operators acting on a
separable Hilbert space. It makes the association
\beq
\Tr\bigl(f_1(y)\cdots f_n(y)\bigr)~\longmapsto~
\frac1{2\pi\,\theta}\,\int\dd^2\mbf r~(f_1\star\cdots\star f_n)(\mbf r) \ ,
\label{WWcorr}\eeq
where the left-hand side of (\ref{WWcorr}) is a trace over
infinite-dimensional operators while the right-hand side is a spatial
integration over functions on $\real^2$. By substituting
(\ref{Xyexpand})--(\ref{WWcorr}) into (\ref{SMCStemporal}) and
(\ref{A0constr}), we arrive at the noncommutative Chern-Simons action
(\ref{SNCStemporal}) with its constraint (\ref{NCGausslaw}). Thus the
matrix model (\ref{SMCSdef}) expanded about the background $X^i=y^i$
as above is exactly equivalent to noncommutative $U(1)$ Chern-Simons
gauge theory. This equivalence is completely analogous to the way in
which noncommutative Yang-Mills theory (\ref{SNCYM}) is derived from
the Yang-Mills potential (\ref{YMpot}) via expansion of matrices about
a noncommutative background.\cite{CDS1,AIIKKT1}

However, as it presently stands, the matrix quantum mechanics expanded
about the noncommutative background describes a system with infinitely many
degrees of freedom. We can remedy the situation, and hence describe a
quantum Hall droplet of {\it finite} size, by introducing a
finite-dimensional version of the matrix Chern-Simons theory
(\ref{SMCSdef}) defined by the matrix-vector $U(N)$ gauged quantum
mechanics with action\cite{Poly1}
\beq
S_N=\frac{m}\theta\,\int\dd t~\Tr\left(\epsilon^{ij}\,X_i\,D_tX_j
  -\frac1{2\,\theta^2}\,X_i^2+2\,A_0\right)+\int\dd t~
\Phi^{I\,\dag}\,D_t\Phi_I \ ,
\label{SNdef}\eeq
where again $X^i$, $i=1,2$ are $N\times N$ time-dependent Hermitian
matrices. The second term in the action (\ref{SNdef}) is a harmonic
oscillator potential for the $X^i$'s, while $\Phi_I$ transforms as a
complex $N$-vector under the gauge group $U(N)$. The crucial point
here is that finite $N$ dimensional representations of the classical
vacua are now possible, with $N=N_e$ identified as the number of electrons
residing in the
lowest Landau level. To see this, we note that the $A_0$ constraint
now selects a sector of particular $\Phi$-charge equal to $m$. We can
solve these constraints classically using the $U(N)$ symmetry of the
model to make $X^1$ diagonal and $\Phi_I$ real. This results in a
system with $N$ {\it real} degrees of freedom and a residual
permutation symmetry generated by the Weyl subgroup of $U(N)$, acting
by permuting the entries of $X^1$ and $\Phi_I$.

In this way, the constrained finite $N$ matrix Chern-Simons theory
(\ref{SNdef}) reduces to the Calogero model for $N$ identical particles moving
in one dimension,\cite{C1,C2} which is defined by the quantum mechanical
Hamiltonian
\beq
H_{\rm C}=\sum_{I=1}^N\left(\frac12\,p_I^2+\frac1{2\,\theta^2}\,
x_I^2\right)+\frac12\,\sum_{I<J}\frac{m(m+1)}{(x_I-x_J)^2} \ .
\label{HCdef}\eeq
A ground state of this Hamiltonian is precisely the Laughlin
wavefunction (\ref{Psi1m}), with $x_I={\rm Re}(z_I)$. One can continue
this and identify all excited Calogero states with excited Laughlin
quasi-particle wavefunctions in a one-to-one manner.\cite{HMVR1} We conclude
that
the finite $N\times N$ matrix Chern-Simons theory is a theory of $N$
composite fermions in the lowest Landau level. The quasi-particles are
well-defined excitations of the noncommutative Chern-Simons gauge
field.

Coming at the noncommutative gauge theory from the string theory side,
it can be shown that certain configurations of D-branes in superstring
theory exhibit the physics of the fractional quantum Hall
effect.\cite{BOB1} The D2-brane effective gauge theory, in a certain
generalization of the Seiberg-Witten scaling limit,\cite{SW1} implies
the effective noncommutative gauge theory described above. In
particular, the role of the electrons in the quantum Hall fluid is
played by D0-branes, and the large $N$ D0-brane matrix model in this way
induces the finite matrix Chern-Simons theory. This provides a string
theory derivation of the proposal that the ground state of the
fractional quantum Hall fluid is described by a noncommutative
Chern-Simons gauge theory.\cite{Suss1} In this way, string theory can
present effective long wavelength descriptions of certain condensed
matter phenomena, and noncommutative field theory provides a surprising
bridge between these two seemingly disparate developments in physics.

\section*{Acknowledgments}

This work was supported in part by an Advanced Fellowship from the
Particle Physics and Astronomy Research Council~(U.K.).

\end{document}